\documentclass[]{aa}
\begin{document}

  \thesaurus{06     
              (12.05.1;  
   	       11.01.2;  
	       11.11.1)}	 
   \title{UV rest frame spectroscopy of four  high redshift ($z>$2)
active galaxies.}

\author{M. Villar-Mart\'\i n,
\inst{1} 
R.A.E. Fosbury,
\inst{2}
 L. Binette,
\inst{3} 
C. N. Tadhunter,
\inst{4}
B. Rocca-Volmerange
\inst{1}}


   \offprints{M.Villar-Mart\'\i n. email: villard@iap.fr}

  \institute{Institute d'Astrophysique de Paris (IAP),
        98 bis Bd Arago, F75014 Paris, France 
	\and
        Space Telescope European Coordinating Facility, Karl Schwarschild Str. 2,
 D-85748, Garching, Germany
        \and
        Instituto de Astronom\'\i a, UNAM, Apartado Postal 70-264, D.F. 04510,
        Mexico
        \and 
	Department of Physics and Astronomy, Sheffield University,
Hicks Building,  Hounsfield
Road, Sheffield S3 7RH, UK
        }

   \date{}

\authorrunning{Villar-Mart\'\i n et al. }
\titlerunning{High redshift active galaxies}

   \maketitle

\begin{abstract}
	
We present UV spectra of a small sample of high redshift
active galaxies: the hyperluminous, gravitationally lensed
 system SMM J02399-0136
($z=$2.8) and three powerful radio galaxies ($z\sim$2.5).
	
	Extended continuum and emission lines are detected in all
objects. The three radio galaxies present typical spectra with
dominant Ly$\alpha$ and weaker CIV$\lambda$1550, HeII$\lambda$1640 and
CIII]$\lambda$1909. The spectrum of SMM J02399-0136 is strikingly
different, showing relatively weak Ly$\alpha$ and HeII and strong
NV$\lambda$1240 relative to the C lines.  

	We find that the weakness of HeII can
be explained if the intermediate density narrow line region ($n\sim$10$^{5-6}$
cm$^{-3}$) dominates the emission line spectrum, rather than the more
extended low density gas ($n\leq$100 cm$^{-3}$). The line ratios of
MRC2025-218 suggest that this is also the case in this object.

	MRC2025-218 and SMM J02399-0136 show the largest NV/HeII and NV/CIV
 values found in high redshift radio galaxies.  The failure of solar 
abundance models
 to explain these line ratios and 
  studies of metal abundances in high redshift quasars and
radio galaxies, suggest that nitrogen is overabundant in the ionized gas 
of these objects.  Alternative possibilities which we discuss include NV
emission from the broad line region and differential amplification 
by a gravitational lens.

	We report the discovery of a P Cygni profile in the CIV
line in the spectrum of MRC2025-218. We detect
also CII$\lambda$1334.5 in absorption and  PCygni profiles for the lines
SiIV$\lambda\lambda$1393.8,1402.8. We do not detect any purely photospheric stellar
lines. The nature of the absorption features is not clear. It could be due
to stars or an associated absorption line system.

\end{abstract}

\section{Introduction}

     High redshift radio galaxies (HzRG) (redshift $z\geq$2)
were until a few years ago the most distant ``galaxies'' (or proto-galaxies)
 we could study. They had a crucial role as probes of the distant universe.
In the late 90's thanks to the Hubble Space  and Keck telescopes, 
many star forming galaxies have been discovered at 
$z\sim$3-4 (see for instance the recent work of Pettini, Steidel, 
Giavalisco). The advantage of these objects is that, unlike HzRG,
the emission is not influenced by the nuclear activity. The study of the stellar component (crucial in Cosmology and galaxy
formation studies)
is in principle easier, since  the emission is dominated  by the stars. 

The very blue continuum observed in HzRG suggested the presence of young 
stars ({\it e.g.} Lilly, Longair \& McLean \cite{lilly83}) and it was believed
that we were observing galaxies in the process of formation.
 However, it was  
discovered later that at least an important contribution to the continuum
radiation is not stellar, but a consequence of the nuclear activity.

One of the most interesting properties of HzRG 
is the so called alignment effect (observed at $z>$0.7):
the con\-tinuum and emission line structures are
 extended  and closely aligned with the
radio axis (Chambers et al.  \cite{chamb87}, McCarthy et al.  
\cite{mac87}). 
The nature of this phenomenon is controversial but it suggests that
the continuum and line emission are closely linked to the nuclear
activity, rather than being due to the stellar population characteristic
of a galaxy in the process of formation.
 Three  main mechanisms have been proposed to explain the alignment 
effect  and all of them could play a role:
1) scattered light from a quasar hidden from the line of sight
(Tadhunter, Fosbury \& di Serego Alighieri \cite{tad89})
 2) young stars whose formation
is triggered by the passage  of the radio jet through the ambient gas (Rees 
\cite{rees89})
3) nebular continuum (Dickson et al.  \cite{dickson95}). The discovery of polarized continuum with the
electric vector perpendicular to the axis of the UV structures provides
strong evidence for the existence of scattered light in many HzRG
at $z>$2 (Cimatti et al.  \cite{cim96}, \cite{cim97}, \cite{cim98}).
Compelling evidence for a young stellar population in a HzRG does not
exist yet, except possibly for  4C41.17 ($z=$3.8) (Dey et al.  
\cite{dey97}). 
In any case, the observed properties of HzRG are a consequence (at least
in the UV rest frame) of the nuclear activity. 

	In spite of this complexity, HzRG still have a crucial role
  in the
understanding of galaxy formation, since they are the only way we have
to study the early stages of {\it giant ellipticals}. Since all powerful
radio galaxies at low redshift are giant ellipticals and there is evidence
that this is also the case at $z\sim$1 (Best, Longair \& R\"ottgering  
\cite{best98}),
it is believed that the host galaxies of radio galaxies at higher redshifts
are also giant ellipticals (Pentericci et al.  \cite{pente99}).   It is however
necessary to  understand how the nuclear activity influences what we
see in order to make a correct interpretation of the observed properties. 

Another interesting aspect is the relationship between the rapid phase(s)
of star formation and the formation and fuel`ing of a massive black hole
during the formation
of these galaxies which are destined to become the massive ellipticals we see
today. Exactly what this relationship is is not clear although 
 interactions/mergers are likely to play a role.

	It has been proposed that ultraluminous infrared galaxies (ULIRGs) 
are progenitors of the giant ellipticals of today (Kormendy \& Sanders,
1992). On the other hand, Sanders et al. (\cite{sand88}) suggested
that ULIRGs will evolve into quasars.  While there is a continuing 
debate about what powers these galaxies
(starburst or active galactic nuclei (AGN))  it is clear that 1) ULIRGS
show clear evi\-dence for  interactions/mergers ({\it e.g.} Borne et al.
 \cite{borne99}, Sanders et al. \cite{sand88}) ~ 
2)  some, at least,  contain powerful AGN  
(e.g. Sanders et al.  \cite{sand88}).

A population  of luminous
galaxies in the submillimetre wavelengths has been discovered in recent years.
 Stu\-dies of the spectral energy
distributions suggest that these are the analogues at high redshift of
ULIRGs  at low redshift ({\it e.g.} Sanders \& Mirabel \cite{sand96}). 
As for many
local ULIRGS, it is not clear whether these galaxies are powe\-red by 
starburst or active galactic nuclei.  The study of distant
ultraluminous submm sources can provide important information about the 
nature of ULIRGs 
and how a massive black hole forms at high redshift and coexists
 with a powerful starburst. The interpretation of the ISM properties
 of these objects in the context of distant radio galaxies then becomes
crucial.

	We study here the UV (rest frame) spectra of three
 HzRG and the hyperluminous submillimetre source SMM J02399-0136
($z=$2.8).

\section{Description of the objects}

	Our sample consists of the following objects:

\begin{itemize}

\item {\it SMM J02399-0136} ($z=$2.80) is described in detail by Ivison  
et al.  (\cite{iv99}). We summarize here the main proper\-ties. This source is
an hyperluminous active galaxy detected in a submm survey with SCUBA. It is
gravitationally lensed by a foreground cluster, which amplifies its luminosity
 by a factor of 2.5. Optical imaging shows two main optical components (named
L1 and L2 by Ivison et al.  \cite{iv98} [IV98 thereafter]) 
separated by
$\sim$ 3 arc sec. 
Both components emit weak continua
and narrow emission lines ($\sim$1000-1500 km s$^{-1}$) that show
 a type 2 AGN in L1 (Seyfert or narrow line quasar).
Radio observations reveal a very weak extended source.
The far infrared (FIR) luminosity is (after correction for
lensing) $\sim$5 times higher than in the
hyperluminous Seyfert 2 FSC10214+4724 (although uncertainties remain about the
exact amplification in this  object). It is not clear whether the bulk of
IR emission is due to dust heated by the active nucleus  or by stars.
If  OB stars are responsible, the FIR luminosity
indicates a star forming rate (SFR) (M$>$10 M$_{\odot}$) of 
$\sim$2000 M$_{\odot}$ yr$^{-1}$.
The source has been also detected in CO showing an  unresolved
($<$5 arc sec) emission spatially coincident with L1.
The gas mass implied by the data is $\sim$10$^{11}$ M$_{\odot}$.
 IV98 propose that this source is associated with a massive starburst.

\item {\it MRC2025-218} ($z=$2.63) was observed by McCarthy et
al. (\cite{mac90}) as 
part of a sample of southern hemis\-phere radio galaxies
selected from the Molongo Refe\-rence Catalogue. 	HST images show that
 the host galaxy has a compact morphology
in the optical (UV rest frame),
consisting of a bright nucleus and two smaller components. Extended
low surface brightness emission elongated and aligned with
the radio axis is also detected. The
galaxy is embedded in a very large halo of ionized gas extended well 
beyond the radio source (Pentericci et al.  \cite{pente99}, 
McCarthy et al.  \cite{mac90}).
Near-IR images (optical rest frame) show that the galaxy
 is quite compact (McCarthy, Perston \& Eisenhardt \cite{mac92},
van Breugel et al.  \cite{breu99}) and fairly symmetric. 

The UV (rest-frame)
spectrum shows strong, spatially extended emission lines (McCarthy 
et al.  \cite{mac90}, 
Villar-Mart\'\i n et al.  \cite{vill99}, VBF99 hereafter),  that 
reveals complex kinematics in
the extended gas. 	Strong continuum emission aligned with
the radio axis is also detected. Its nature is uncertain but its
high level of polarization (Cimatti et al.  \cite{cim94}) shows that
the contribution of a scattered component is important.
The authors found that 
the rest frame UV continuum emission is linearly polarized 
($\sim$8.3$\pm$2.3\%) with the
electric vector perpendicular to the UV emission axis. Any valid model
for the spectral energy distribution   (SED) requires at least
 two  components: a polarized
scattered component in the UV and a redder, probably unpolari\-zed
component, best represented by an evolved stellar population with a
minimum age of about 2 Gyr. A nebular component associated with
the line emitting gas is also required.

\item {\it MRC1558-003} ($z=$2.53) was part of the sample of ultra steep
spectrum (USS) radio sources of R\"ottgering et al.  (\cite{rott95}). R band
CCD images  show a rather small source extended ($\sim$1.8 arc sec is
the largest extension) along position angle (PA) 50\degr. The radio
axis PA is 75\degr ~(R\"ottgering et al. \cite{rott94}). 
UV (rest frame) spectroscopy shows strong Ly$\alpha$ emission along the
radio axis direction 
with a bright component and more diffu\-se emission
 extending for at least 15 arc sec. CIV is also extended.
The lines show complex kinematics in the extended gas
with FWHM$\sim$1000-1500 km s$^{-1}$ (VMBF99). 

\item {\it MRC2104-242} ($z=$2.49) was also observed by McCarthy et al. 
(\cite{mac90}) 
selected from the Molongo Refe\-rence Catalogue. 
Broad band images (McCarthy et al.  \cite{mac90}) show
 two bright clumps between the radio lobes and aligned with
them. Narrow band Ly$\alpha$  images show emission extended over more
than 15 arc sec along the radio axis. A large halo of diffuse emission
seems to surround the entire object.
The two main blobs emit continuum and strong emission lines.
 UV (rest frame) spectroscopy reveals complex kinematics with FWHM$\sim$1500
km s$^{-1}$ (McCarthy, Baum \& Spinrad \cite{mac96}, VMBF99). HST images 
(Pentericci et al.  \cite{pente99})
show that the host galaxy is very clumpy. There is also a narrow filament 
extending for $\sim$2 arc sec and aligned with the radio axis within a few
degrees.

\end{itemize}

\section{Observations and Data Reduction}

The spectroscopic observations were carried out on the nights 1997 July 3-5
and 1998 July 25-27
using  the EMMI multi-purpose instrument at the NTT (New Technology
Telescope) in La Silla Observatory
(ESO-Chile). The observations are described in detail in VMBF99.

	The data reduction was done  using standard methods in IRAF. The 
spectra  were bias subtracted and divided by a normalized 
flat-field frame (dome 
flat-field). Illumination corrections along the slit were found to be
ne\-gligible. The spectra were calibrated in wavelength using comparison spectra
of HeAr. Cosmic rays were removed automati\-cally (we obtained at least
three similar frames for each object).
Sky lines were subtracted and the spectra correc\-ted for atmospheric
extinction with the aid of mean extinction coefficients for La Silla.
For each night we built a mean response curve from the standard stars
observed that night with a wide slit (5 arc sec). Each object frame was flux
calibrated with the corresponding response curve.  The spectra were also
corrected for galactic reddening. The reddening values were based on
Burstein and Heiles  (\cite{burs84}) maps, using empirical selective function of
Cardelli, Clayton \& Mathis (\cite{card89}). The redshifted CIII]$\lambda$1909 for SMM J02399-0136
is  contaminated
by the atmospheric absorption band at $\sim$7250 \AA. We created the
spectrum of the atmospheric band from 
the 1-D spectrum of a standard star taken with the same slit width as the one
used for the radio galaxy (1.5 arc sec). We fitted the continuum and divided
the original spectrum by the fit. By removing the absorption features intrinsic
to the star,
we obtained the spectrum of the atmospheric band. We divided the spectrum
of SMM J02399-0136 by it trying different factors.
Our conclusion is that the remaining  effect of the atmospheric absorption is
negligible after correction. 

	Both IRAF and STARLINK (DIPSO) ~routines were used to measure the
emission line fluxes, FWHMs and wavelengths.  1-D spectra were extracted
for all objects from  apertures described below. 
We then fitted Gaussian profiles 
to the lines. The measured
flux and FWHM of the spectral line are the values measured from the fit of 
 the Gaussian. 
The instrumental profile  was subtracted
from the observed FWHM in quadrature.

\section{Results}

\begin{figure*}[ht]
\includegraphics{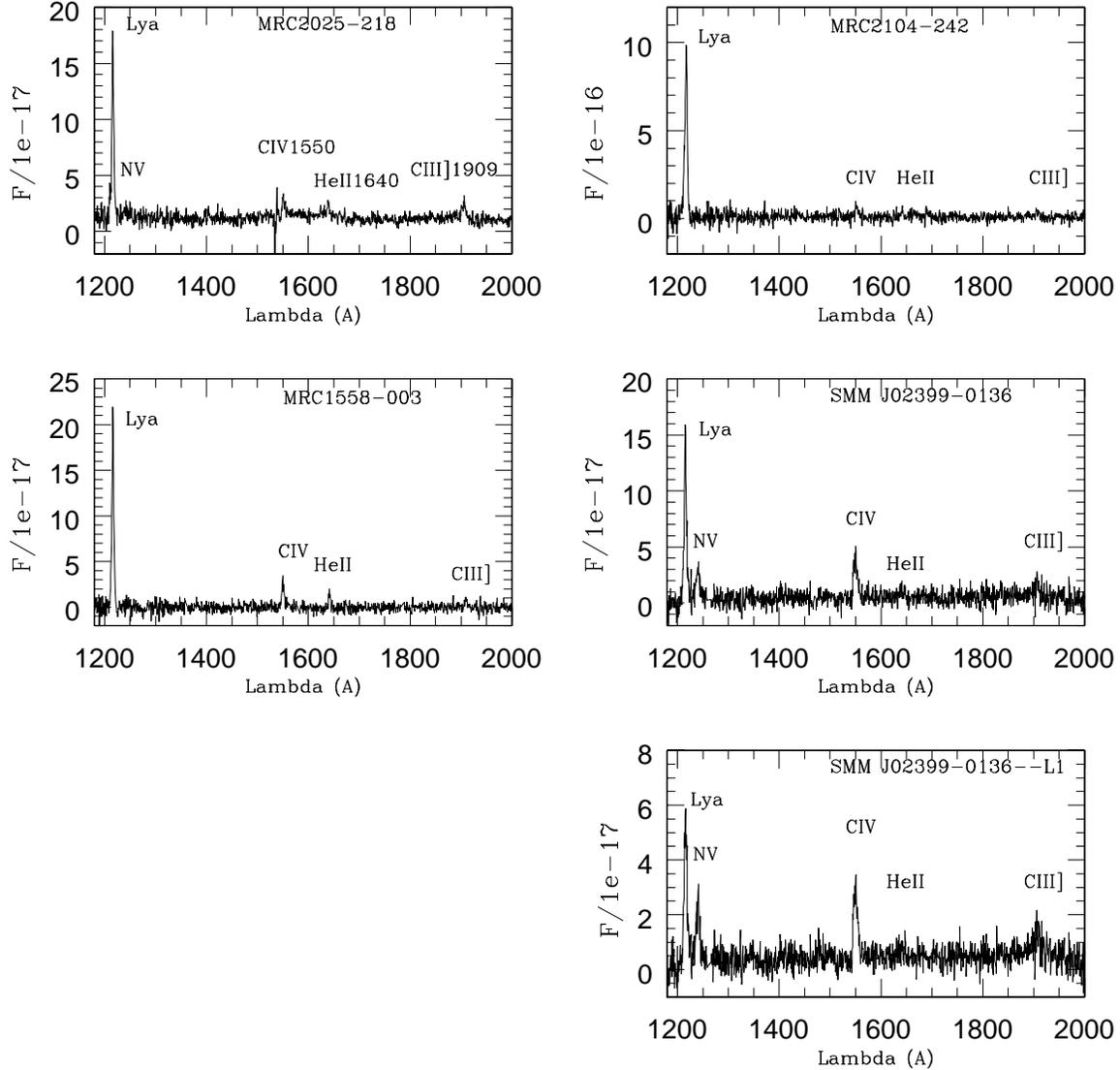}
\vspace{6.in}
\caption{Integrated spectra of the four objects in the sample discussed
in this paper. The last spectrum corresponds to component L1 (adopting
IV98 nomenclature) in the
system SMM J02399-0136 (the active galaxy). Flux is given in units
of 10$^{-17}$erg s$^{-1}$ cm$^{-2}$ \AA$^{-1}$, except for MRC2104-242, which
is given in units of 10$^{-16}$erg s$^{-1}$ cm$^{-2}$ \AA$^{-1}$.}
\label{Fig1}
\end{figure*} 

\subsection{The spectra}

The spatially integrated
spectra of our objects are shown in Fig. \ref{Fig1}. We extracted each spectrum
from an  aperture  inside which the weakest lines were detected with
higher S/N: MRC2025-218 (1.9 arc sec, centered at the spatial
continuum centroid); MRC2104-242 (5.9 arc sec centered at 
the intersection between the two bright clumps); MRC1558-003 (2.7 arc sec
centered at the spatial position of centroid of the brightest Ly$\alpha$
component).
 The two spectra of
 SMM J02399-0136  correspond to a) the whole system (L1+L2) 
(the spectrum was extracted form a 5.4 arc sec aperture covering the brightest
emission of the L1+L2 components) and b)  he AGN component, L1. 
(2.2 arc sec aperture centered at the
centroid of Ly$\alpha$ emission in L1)

\begin{table*}
\normalsize
\centering
\small
\caption{CIV flux (in units of 10$^{-16}$ erg s$^{-1}$ cm$^{-2}$. The fluxes
of the strongest lines are given  relative to CIV. The FWHM of the lines are also shown}
\begin{tabular}{llllllll}
	&	&~~~~Line	&ratios	&	&	\\ \hline
Target	& F${_{CIV}}_{16}$   &  Ly$\alpha$/CIV   & NV/CIV &  HeII/CIV & CIII]/CIV  \\ \hline
MRC2025-218 	& 0.69$\pm$0.07	& 5.7$\pm$0.9  	&  0.9$\pm$0.3	& 0.5$\pm$0.2	& 1.4$\pm$0.3  	\\
MRC2104-242	& 3.8$\pm$0.7	& 15$\pm$3 	&  $\leq$1 &	 0.5$\pm$0.2	& 0.7$\pm$0.3  \\
MRC1558-003	& 1.7$\pm$0.1	& 5.4$\pm$0.4	&  $\leq$0.35 &	 0.43$\pm$0.08	& 0.36$\pm$0.09  \\
SMM J02399-0136	& 3.3$\pm$0.2	& 2.5$\pm$0.2	&  0.8$\pm$0.1 & 0.26$\pm$0.05	& 0.42$\pm$0.07	  \\ 
SMM J02399-0136$_{L1}$ & 2.5$\pm$0.1 & 1.6$\pm$0.1	&  0.9$\pm$0.1 & 0.20$\pm$0.07 & 1.0$\pm$0.2  	\\ 
	&	&	FWHM	& (km s$^{-1}$) &	&	\\ \hline
Target	& Ly$\alpha$   &  NV &   CIV   & HeII & CIII] \\ \hline
MRC2025-218 	& 700$\pm$100  & 2800$\pm$800	& 600$\pm$200$^a$	& 600$\pm$300	&
1400$\pm$200	\\
MRC2104-242	& 1300$\pm$100  & 	& 1000$\pm$400	& $<500$	& 1000$\pm$300	\\
MRC1558-003	& 800$\pm$100  & 	& 1000$\pm$150	& 500$\pm$200	&  850$\pm$300	\\
SMM J02399-0136	& 1200$\pm$100 & 2000$\pm$300	& 1600$\pm$100	& 1800$\pm$600	& 1100$\pm$700	\\ 
SMM J02399-0136$_{L1}$	& 1600$\pm$100	  & 2000$\pm$300 & 1600$\pm$100	& 2100$\pm$600		& 6100$\pm$1200		&	\\ \hline
\end{tabular}
\begin{tabular}{l}
$^a$ The absorption on the blue side of the line (see below) has not been taken into
account
\end{tabular}
\end{table*}

\begin{figure}
\includegraphics{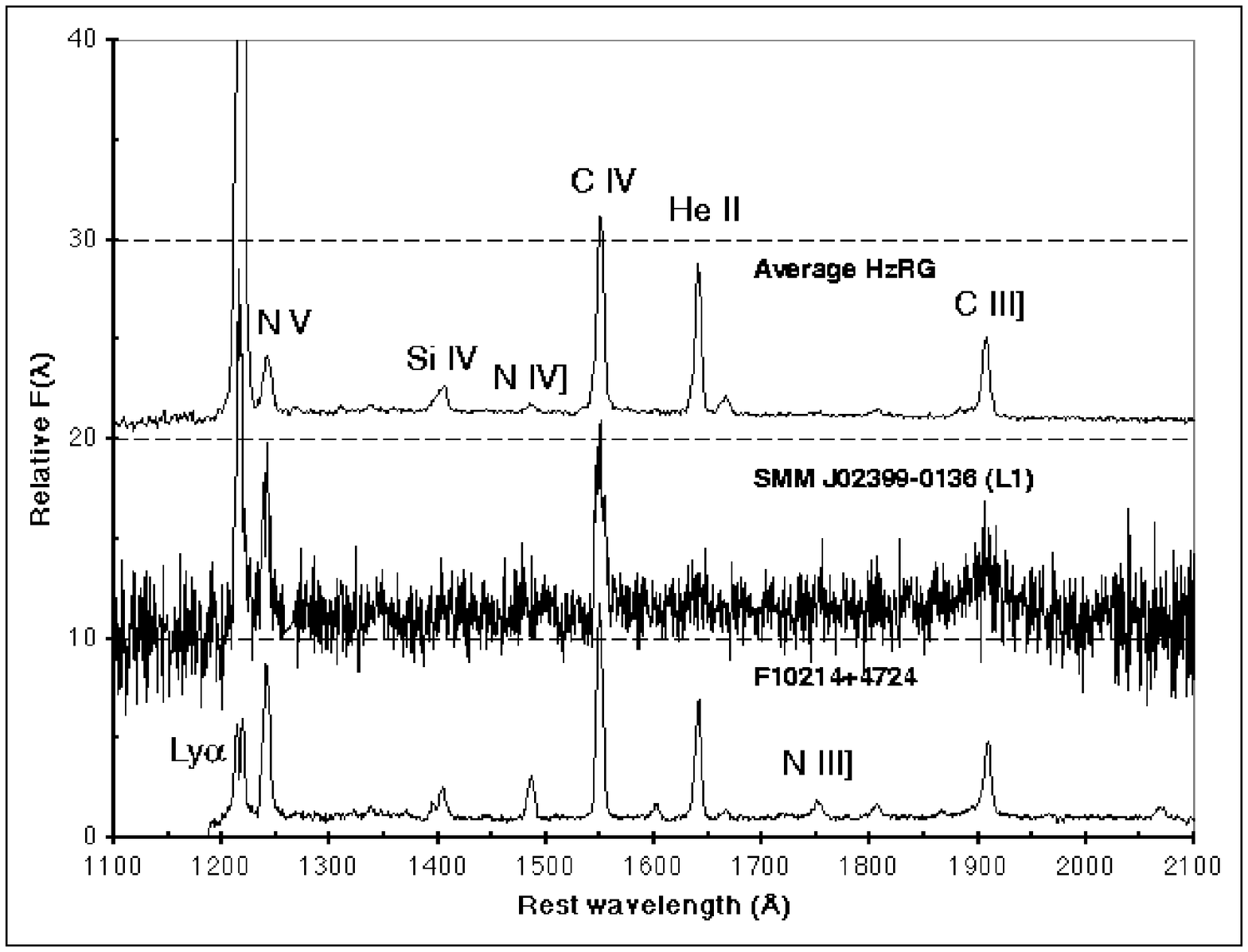}
\vspace{3.in}
\caption{Comparison between the spectrum of L1 (the AGN in SMM J02399-0136),
an average HzRG spectrum (Vernet et al. \cite{ver99}) and the hyperluminous Seyfert 
2 galaxy FSC10214+4724 (Goodrich et al.  \cite{goo96}). Notice the
relative weakness of Ly$\alpha$ and HeII and the strength of NV in L1 
compared to a typical spectrum of HzRG.}
\label{Fig2}
\end{figure}

The spectra (except for SMM J02399-0136) are  typical of HzRG.
Ly$\alpha$ is the strongest line and  weaker
CIV$\lambda$1550, HeII$\lambda$1640 and 
CIII]$\lambda$1909  are also detected. 
NV$\lambda$1240  is detected only in MRC2025-218 and
SMM J02399-0136. This line is often not detected in HzRG (R\"ottgering
et al.  1996). We compare in  Fig. \ref{Fig2}  the spectrum of L1
with an average spectrum of HzRG (Vernet et al. \cite{ver99}) and the hyperluminous
(also gravitational lensed) Seyfert 2 galaxy FSC10214+4724
(Goodrich et al.  \cite{goo96}). The differences
are striking. L1 presents very weak HeII, strong NV and weak Ly$\alpha$
compared to  typical HzRG spectra. It is similar to FSC10214+4724 in the sense
that Ly$\alpha$ is weak and NV strong, but HeII is relatively much weaker
in L1.

	We present in Table 1 some parameters characterizing 
the main UV emission lines, resulting from 1-D Gaussian fits
to the line profiles.

\subsection{The spatial distribution of the continuum and the emission
lines}

\begin{figure}
\includegraphics{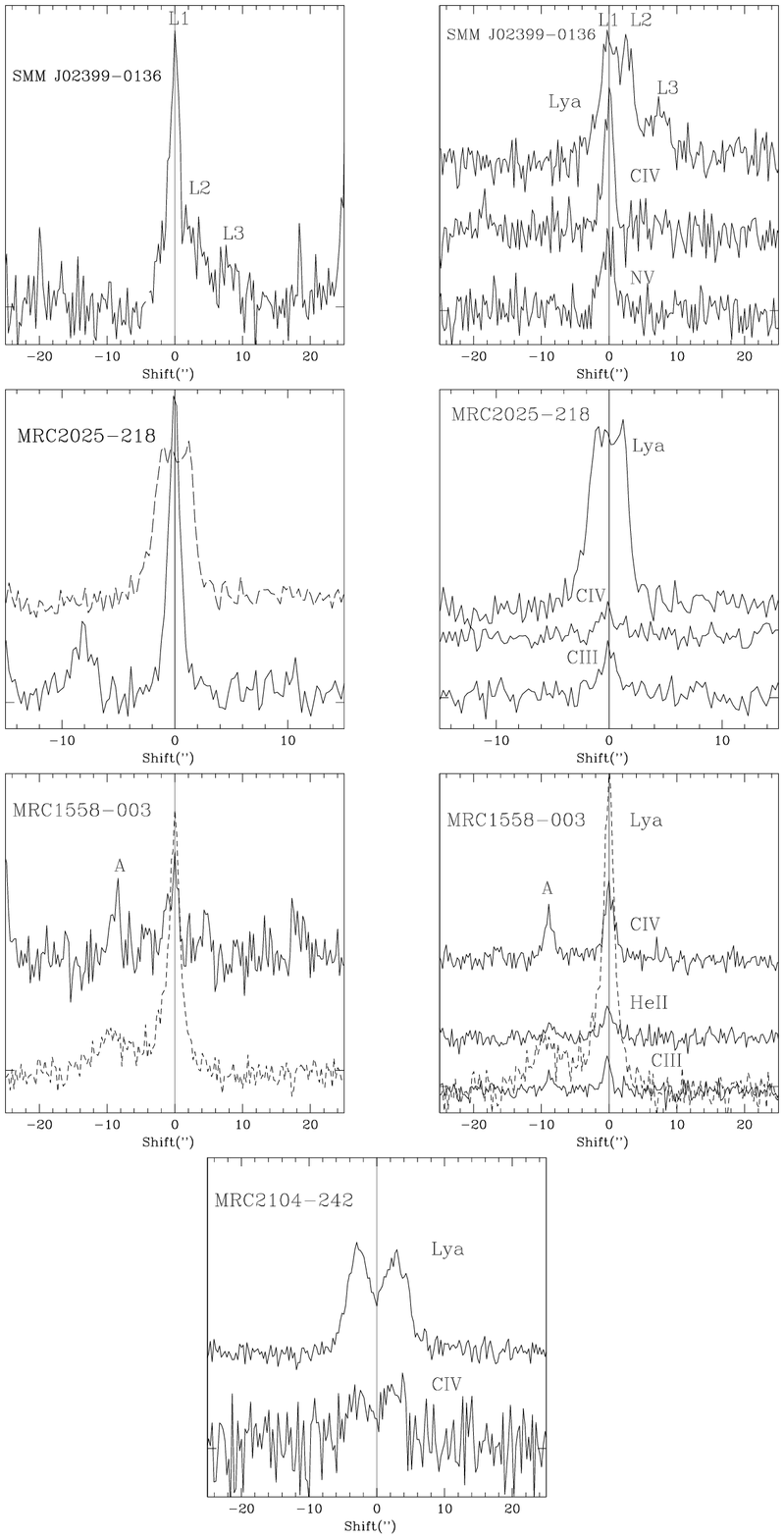}
\vspace{6.7in}
\caption{Spatial profiles of the continuum and brightests UV lines.
Left panels: Continuum in solid lines and Ly$\alpha$ in dashed lines (when shown).
Right panels: Emission lines only. Bottom: Emission line profiles in
MRC2104-242. In all panels a constant has been added to some of the
profiles to make the plots clearer. The spatial zero for each object
 has been
selected as the spatial position of the continuum centroid, except
for MRC2104-214 that has a very weak continuum. For this object
we selected the spatial position which appears to
best define the separation between the two main optical components. }
\label{Fig3}
\end{figure}

	 We described in VMFB99 the spatial properties of the Ly$\alpha$
emitting gas derived from the 2-D spectra. Here we present 1-D spatial
profiles for Ly$\alpha$, the continuum and the strongest UV lines.
We created the emission line spatial profiles by adding those pixels
(in the dispersion direction) where a given line is detected and 
subtracting the underlying continuum 
 from a window of identical size (in \AA )
  close to the line. The continuum spatial profile was created using
a much larger window to increase the S/N ratio 
(typically 100-150 pixels in the dispersion direction, {\it i.e.} $\sim$
65-90 \AA ). We present in Fig. \ref{Fig3} the spatial profiles for
the strongest lines and the continuum.

\begin{itemize}

\item {\it SMM J02399-0136:} The two optical components (L1 and L2) 
identified by IV98
are clearly seen both in continuum (dominated by L1) and Ly$\alpha$
(similar strength in both components) separated by $\sim$ 3 arc sec. 
L1 is detected in the other
UV lines as well, but not L2 (the {\it spatially integrated} spectrum  of
L2 shows weak CIV).  Both the Ly$\alpha$ and the continuum  profiles
reveal a region  beyond L2 (named L3 in the figures) up to $\sim$10 arc sec 
beyond
the continuum centroid of L1.  VMBF99 showed that Ly$\alpha$ presents
large velocity widths ($\sim$1000 km s$^{-1}$) in L3. 
No other lines are detected in
this region except possibly CIV.  L1 is rather 
compact, but it appears to be marginally resolved
 (1.45 arc sec, while
the average seeing was 1.20 arc sec). 
It contains an unresolved
component (WF/PC1 F702W and WFPC2 F336W images 
 show an unresolved source at the L1 position 
with a FWHM of $\sim$0.1-0.2 arc sec [IV98]) and some extended emission.

\item  {\it MRC2025-218:}  
 The  spatial profiles of  Ly$\alpha$ 
and the  continuum reveal very different distributions. The con\-tinuum
emission is extended, but the dominant component is just marginally resolved
(1.3 arc sec while the average seeing was 1.1).
Ly$\alpha$ is more extended and the flux does not peak at the 
position of the con\-tinuum centroid, but it presents a plateau.
CIV (maybe CIII] as well) are also  extended. The 2-D spectrum (see VMBF99)
reveals a bimodal distribution for Ly$\alpha$ which was also mentioned
by McCarthy et al. \cite{mac90}

\item {\it MRC1558-003:} The Ly$\alpha$ profile reveals the presence of
two main components separated by $\sim$9 arc sec: the main optical
component and a region named A in the Figure. VMBF99 showed that this
A region emits lines of large widths (FWHM$\sim$1500 km s$^{-1}$). Both
components emit  also CIV, CIII], 
HeII  and continuum. The continuum and the emission lines (at least CIV and
Ly$\alpha$) are spatially resolved in both regions. 

\item {\it MRC2104-242:} 
Ly$\alpha$  presents a bimodal distribution, clearly seen in CIV
as well. Both components are spatially resolved. The centroids are separated
by $\sim$7 arc sec. We do not present the spatial profile of the continuum 
in Fig. \ref{Fig3}  because it is too noisy, but very weak continuum is detected  
associated with the two Ly$\alpha$ components. 

\end{itemize}

	Therefore, all objects present extended continuum  
  over several
arc sec (up to $\sim$15 arc sec in SMM J02399-0136). The continuum in
MRC2025-218, MRC1558-003 and SMM J02399-0136 is dominated by a bright 
component which is
rather compact. It is probably unresolved in  SMM J02399-0136 and marginally 
resolved in
 MRC2025-218. All objects show
extended emission lines which present rather different spatial profiles
compared to the continuum (compare, for instance, 
the Ly$\alpha$ and continuum profiles in SMM J02399-0136 and MRC2025-218).
Both lines and continuum reveal the presence of several spatially distinct 
regions.

\subsection{Absorption features in the spectrum of MRC2025-218}

\begin{figure}
\includegraphics{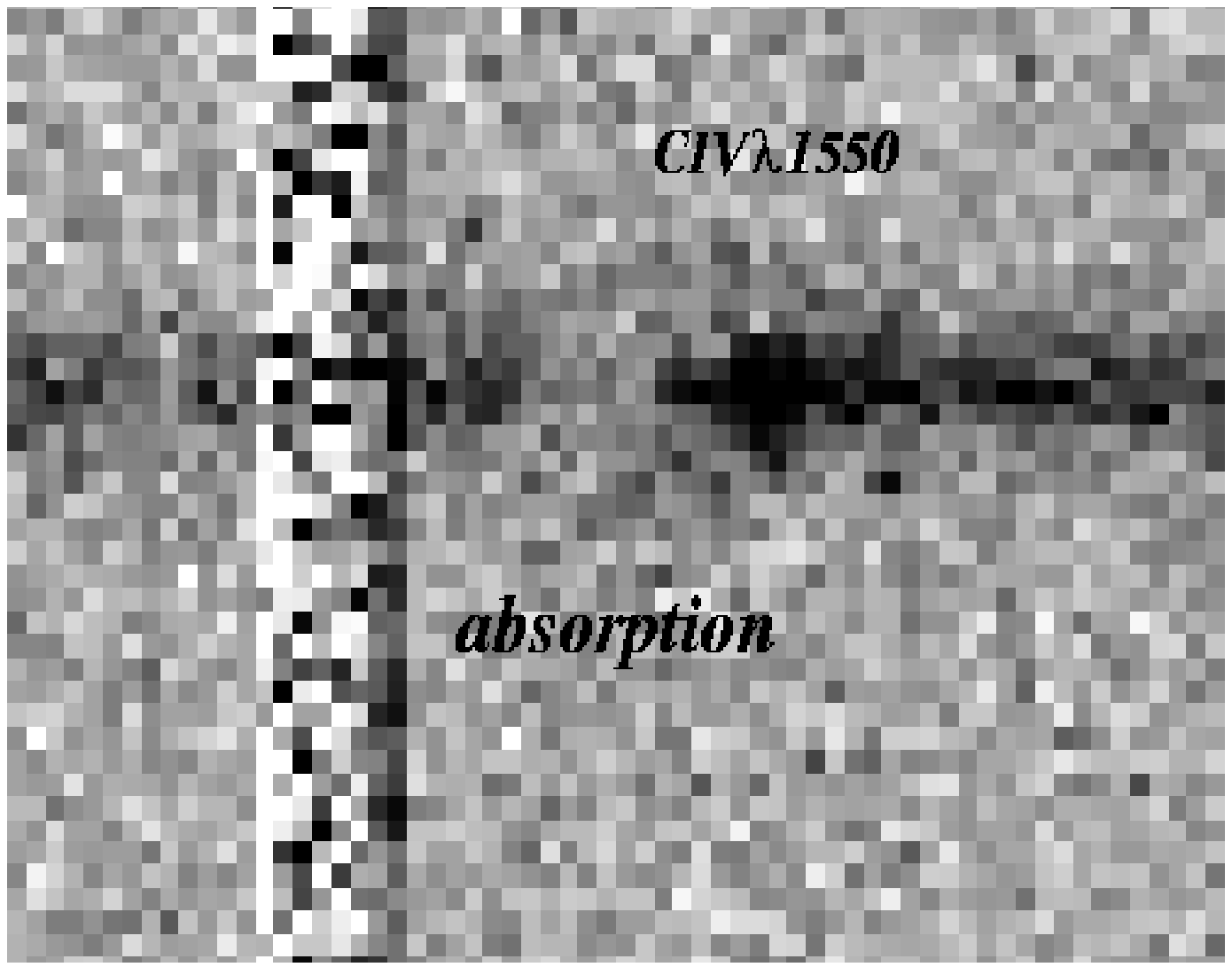} 
\vspace{2in}
\caption{2-D spectrum of MRC2025-218 in the CIV region. There are some 
residuals of a night sky line on the blue side clearly seen in the figure.
Spectral dispersion runs in the horizontal coordinates and spatial direction in vertical coordinates.}
\label{Fig4}
\end{figure}

\begin{figure*}
\includegraphics{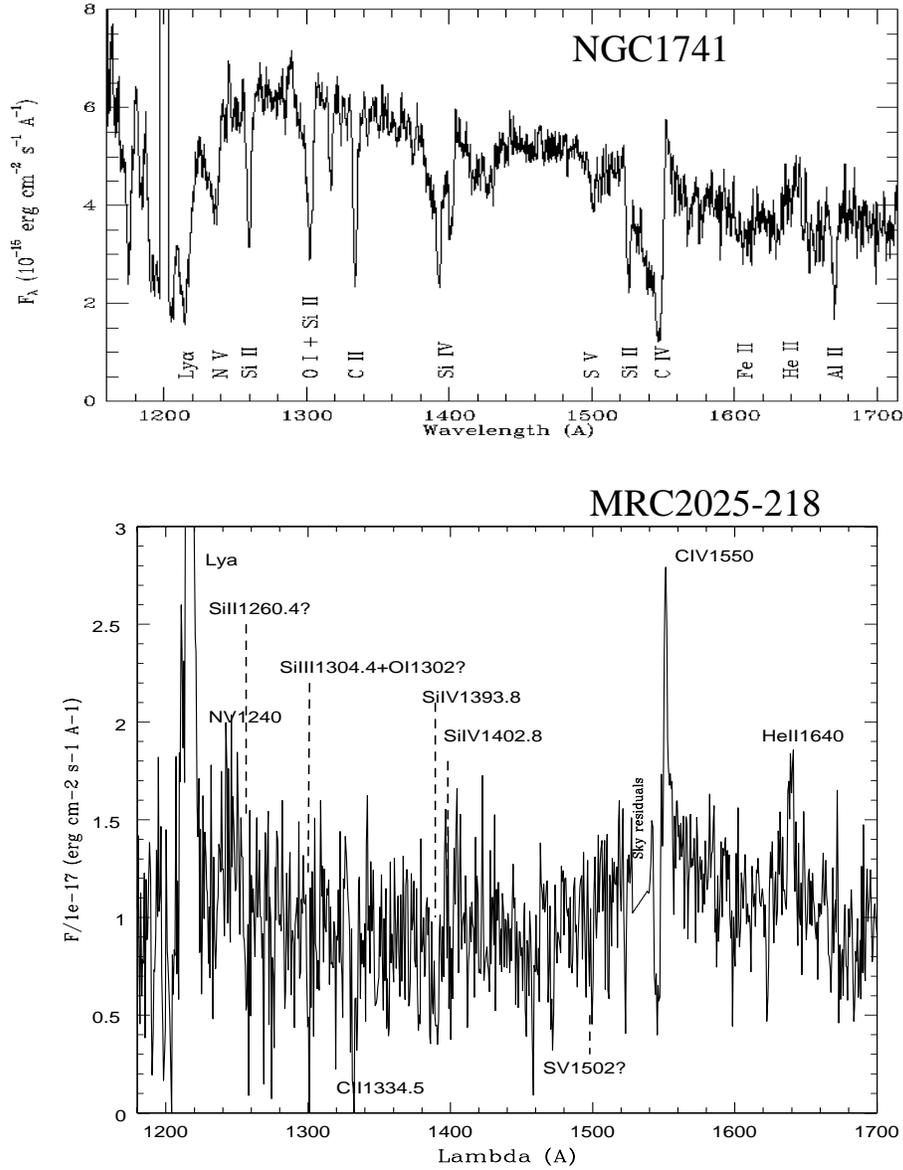} 
\vspace{6.1in}
\caption{Top: Spectrum of the star forming knot B1 in the nearby starburst
 galaxy NGC1741 (Conti, Leitherer \& Vacca \cite{conti96}).
Bottom: Spectrum in the same spectral region of MRC2025-218. We indicate in this diagram the possible absorption features
detected in MRC2025-218.}
\label{Fig5}
\end{figure*}

	The 2-D spectrum of MRC2025-218 shows a clear absorption
feature blueshifted with respect the CIV emission (see
Fig. \ref{Fig4}). In order to  search for other absorption features, we have
extracted a 1-D spectrum from the continuum emitting
region  (8 pixels or 2.2 arc sec aperture). 
We fitted the profiles of all possible absorption detections. 
We assumed Gaussian profiles (although it does not necessarily have
 to be the case).
Some absorption features are detected. We show in Fig. \ref{Fig5} (bottom) 
the spectrum
in the range  1180-1700 \AA , with the expected position of some 
absorption features commonly found in nearby starburst galaxies. 
We present for comparison the spectrum of the B1 star forming
 knot in NGC1741 (Fig. \ref{Fig5} top)
(Conti, Leitherer \& Vacca \cite{conti96}).
	
	We rejected those features such that:
a) the spectral profile was (taking errors into account) narrower than the
 instrumental profile (IP)
(2.98 \AA\ in the rest frame) and/or b) the absorbed flux was lower than the detection limit.
This was the case of SiII$\lambda$1260.4, OV$\lambda$1371, SiIII$\lambda$1417, SV$\lambda$1502.
In order to calculate the detection limit for an absorption feature at a given position, we
created Gaussians with the expected FWHM (IP in all cases) and varied the amplitude (the profiles 
could be broader, but this just means that  a larger flux
would be needed for detection). The Gaussians
were added to the continuum near the expected  position. The upper
limit was chosen by eye, as the flux of that Gaussian  that we considered detectable.

\begin{figure}
\includegraphics{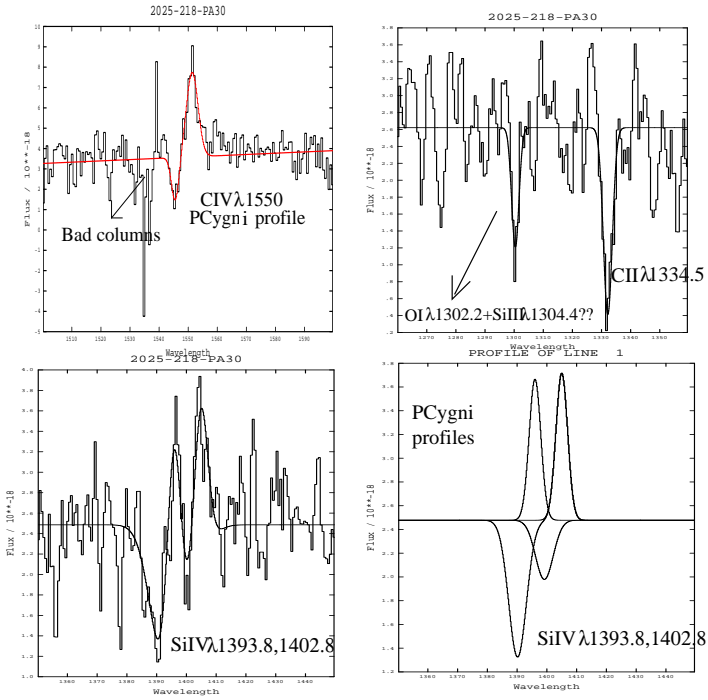} 
\vspace{4in}
\caption{Absorption lines in the spectrum of MRC2025-218. The fit and the data
are presented for all lines.   Note the P Cygni profile in CIV and  SiIV$\lambda\lambda$1393.8,1402.8.
The panel on the right-bottom side shows the individual components fitted to the SiIV lines.}
\label{Fig6}
\end{figure}

	We present in Fig. \ref{Fig6} the fits to those features that we accepted as real. 
Except for CIV (for which the original spectrum is shown), we present smoothed spectra
to make the figures clearer. The fits were done to the original (non-smoothed) spectra.
There are some sky emission residuals on the
blue side of the CIV absorption feature, but they do not affect the fit 
(see Fig. \ref{Fig4}). 
We present in Table 2 some parameters obtained from the fits:
wavelength, line identification, EW (rest frame) and FWHM.  The values 
measured in the radio galaxy 4C41.17 ($z=$3.80) (Dey et al.  1997) and
the associated absorption system of the quasar 3C191 ($z=$1.95) 
(Bahcall,  Sargent \& Schmidt 1967 \cite{bahc67})  are also shown.

 \begin{table*}
\small
\centering
\caption{Absorption lines measured in the spectrum of MRC2025-218. The EWs
are given in the rest frame. Parameters
of the emission line are given for those cases where a PCygni profile is detected. We present also 
the values measured  for 4C41.17 (Dey et al. \cite{dey97}) and 3C191 
(Bahcall, Sargent \& Schmidt \cite{bahc67}).}
\begin{tabular}{lllllllll}
\hline
 $\lambda$	& Line	        & EW (\AA )	& EW  	       & 	EW           & FWHM  km s$^{-1}$      & FWHM	     \\ 
		&  	&  $_{2025-218}$	& $_{4C41.17}$ & 	$_{3C191}$   & $_{2025-218}$ & $_{4C41.17}$ &	\\ \hline
1300.4$\pm$0.5	&  OI$\lambda$1302.2$_{abs}$ &   1.7$\pm$0.7 &  1.4$\pm$0.3	&  & $<$690  	& 640$\pm$160 & 	\\
		&   +SiII$\lambda$1304.4			&		&	&		&		&		\\
1332.2$\pm$0.5	&  CII$\lambda$1334.5$_{abs}$	& 3.2$\pm$0.6 &	1.15$\pm$0.25  & 3.39	&  500$\pm$200	& 950$\pm$220 	\\
1390.145$\pm$1	&  SiIV$\lambda$1393.8$_{abs}$	&   2.9$\pm$0.6	&  1.15$\pm$0.25 &   3.39  &  1400$\pm$500	&  445$\pm$70 & \\
1395.996$\pm$0.6 &  SiIV$\lambda$1393.8$_{emis}$ &   1.3$\pm$0.6	&  1.31 &  	  & 800$\pm$300 &  1130$\pm$170 \\
1399.145$\pm$1 	&  SiIV$\lambda$1402.8$_{abs}$	&  2.5$\pm$2 	& 0.5$\pm$0.2   &  2.38	  & 1400$\pm$500	& 362$\pm$62 \\
1404.996$\pm$0.6	&  SiIV$\lambda$1402.8$_{emis}$	&  4.0$\pm$2.5 	& 1.75    &   & 800$\pm$300 & 1130$\pm$170	\\ 
1545.5$\pm$0.5	& CIV$\lambda$1550$_{abs}$	& 10$\pm$3	& 1.8$\pm$1.3   &   7.13	& 400$\pm$200 	&748$\pm$210$^a$ \\
 1551.4$\pm$0.3	& CIV$\lambda$1550$_{emis}$	&  17$\pm$4 	&  21   & 	& 700$\pm$200  & 540$\pm$15$^a$ \\
\hline
\end{tabular}
\begin{tabular}{l}
$^a$ FWHM of the individual components in the CIV doublet. \\
\end{tabular}
\end{table*}

	The fitting to  SiIV$\lambda\lambda$1393,1402 is difficult due to the 
low S/N ratio, however, the presence of two P-Cygni profiles
for the SiIV$\lambda$1393.8 and SiIV$\lambda$1402.8 lines is sugges\-ted by the data. The best fit
is obtained by considering two absorption and two emission features.
We constrain the fit so that the two absorption features on one hand, and
the two emission features on the other, have the same FWHM and are separated
by 9 \AA ~(as is expected from the doublet components
$\lambda\lambda$1393,1402). The results of the fit are presented in Table 2. The FWHM of the 
emission line
features are consistent with the value measured for the CIV and Ly$\alpha$ emission.
The absorption feature presents a much larger FWHM than measured for CIV absorption. 
The value is consistent with measurements
in galactic outflows.

	The absorption features are narrower than the emission lines (also narrower than in 4C41.17) 
except for the SiIV$\lambda\lambda$1393,1402  
lines, that have FWHM$_{abs}\sim$1400$\pm$500. The absorption  features in the P Cygni profiles are
blueshifted
 by 1100$\pm$200 km s$^{-1}$ (CIV) and 1200$\pm$200 (SiIV$\lambda$1393.8 and 
SiIV$\lambda$1402.8) with respect to the emission.

	We confirm the detection in absorption of CIV$\lambda$1550, CII$\lambda$1334.5, SiIV$\lambda\lambda$1393.8 +1402.8
and, maybe, OI$\lambda$1302.2 +SiII$\lambda$1304.4 (the absorbed flux is slightly higher than the detection
limit). 	There is no clear evidence in our data
for Ly$\alpha$ absorption, although the asymmetry of the profile  (steeper on the blue side)
 is probably 
due to absorption. 
The presence of P Cygni profiles is confirmed in the case of CIV,
 SiIV$\lambda$1393.8 and SiIV$\lambda$1402.8.

\section{Discussion}

\subsection{The spatial distribution of the emission lines and continuum.}

	All the objects in the sample present extended continuum and
line emission. These structures are aligned with the radio axis 
at least in MRC2025-218 and MRC2104-242 (McCarthy et al.  1990, Pentericci
et al.  1999). L1 and L2 in SMM J02399-0136 define a line with position angle 
88\degr while the radio axis position angle is 71\degr. Therefore, the optical and
radio axis are closely aligned in this object as well.
 We located the slit in MRC1558-003 aligned with
the radio axis and therefore, strong line and continuum emission is 
extended in this direction. 

	Our spectra reveal the presence of several spatial components 
in all objects.  A clumpy morphology has been observed in most HzRG
 both in continuum and Ly$\alpha$ 
({\it eg.} Pentericci et al.  1999).  According to the properties of the individual clumps,
Pentericci et al.  have  suggested that we are witnessing the
merging of  several sub-units to form the host galaxy of the radio
source.

\subsection{The emission line ratios.}

\begin{figure*}
\includegraphics{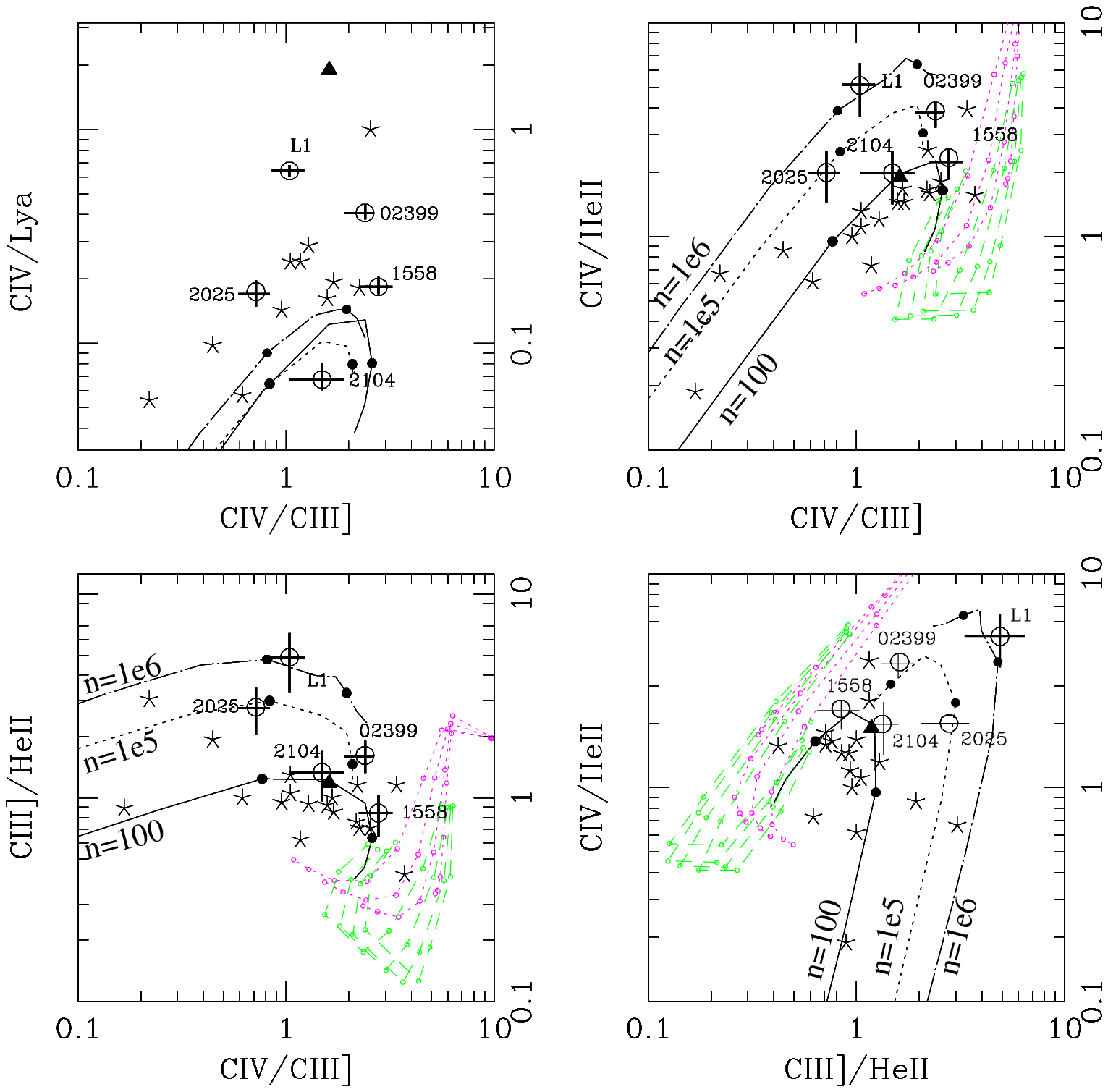}
\vspace{5in}
\caption{Diagnostic diagrams involving the strongest UV lines. Our objects are represented as
open circles. Stars are HzRG in van Ojik's (\cite{ojik95}) sample. 
The solid triangle is the hyperluminous
radio galaxy FSC10214+4724 (data obtained from Goodrich et al.  \cite{goo96}). 
The solid line is an AGN
photoionization sequence (in $U$, the ionization parameter) 
with a power law of index $\alpha$=-1.0, solar abundances and
low density ($n=$100 cm$^{-3}$).  Similar sequences with $n=$10$^5$ and  
$n=$10$^6$ cm$^{-3}$
 are shown.   The solid circles in each of these three sequences show
the models with U=0.01 and 0.1. Notice that MRC2025-218 and L1 are better
explained by the higher density models, with U$\sim$0.01.
The shock models (dotted lines)
and shock+precursor models (dashed lines) from Dopita \& Sutherland (\cite{dop96}) are presented as well. 
 The weakness of Ly$\alpha$ in the first diagram is probably due to
absorption by neutral H possibly combined with dust.}
\label{Fig7}
\end{figure*}

We have compared the line ratios of the objects in our  sample
with measurements 
for other HzRG (van Ojik et al. \cite{ojik95}) by plotting them 
in four diagnostic diagrams invol\-ving the
main UV lines (Fig. \ref{Fig7}).
We plot also some models predicting the UV line ratios of HzRG when
a) AGN is  the dominant ionization mechanism b) shocks dominate the emission
line processes (see Villar-Mart\'\i n et al.  1997 for a detailed 
discussion of the models). The AGN photoionization models 
were built with the photoionization code
 MAPPINGS~Ic (Luc Binette), which  is described in  Ferruit et al. (\cite{ferr97}). 
The shock models are taken from Dopita \& Sutherland (1996).
Villar-Mart\'\i n et al. (\cite{vill97}) 
showed that most HzRG define a sequence which can be
explained in terms of  (AGN) photoionization by a power law of index
$\alpha$=-1 (with $F_{\nu}\propto \nu^{\alpha}$) of a low density gas
($n\leq$100
cm$^{-3}$) with solar abundances. The parameter sequence is the ionization parameter U, so
that the di\-fference from object to object is due to a difference in
the ionization level of the gas, produced either by geometric dilution
or by differences in the quasar luminosity.

	MRC2104-242 and MRC1558-003 lie in the general sequence 
defined by most HzRG.
This is not the case for the integrated spectrum (L1+L2) of
the system SMM J02399-0136, which presents  weak Ly$\alpha$ 
and weak HeII relative to CIV and CIII]\footnote{The CIII] profile is 
contaminated in quasars by  SiIII]1895. This could also be the case in L1 and
therefore, the real value of the CIII] line  ratios is uncertain.}
   This behaviour is more pronounced
if we extract
the spectrum of the L1 component (the active galaxy).  The line ratios 
locate the object far beyond the general
trend, due to the weakness of HeII and Ly$\alpha$ (see the spectrum in Fig. \ref{Fig2}).

	MRC2025-218 lies also far from the general trend.  
 Since CIV/HeII is similar to the values observed in other HzRG 
and  CIV/CIII] and CIII]/HeII are too  low,  the discrepancy is probably due
to the unusual strength of CIII] in this source, rather than HeII being too weak, as it
might seem from  the diagnostic diagrams .\footnote{We detect  SiIII]1895 
in MRC2025-218, but they are well resolved from CIII] and are not likely
to contaminate the profile of this line. On the other hand, CIV is absorbed 
in this object [see previous section], but according
to the fit  shown in Fig. \ref{Fig6}, the absorbed flux is only a few per cent)}

\begin{figure}
\includegraphics{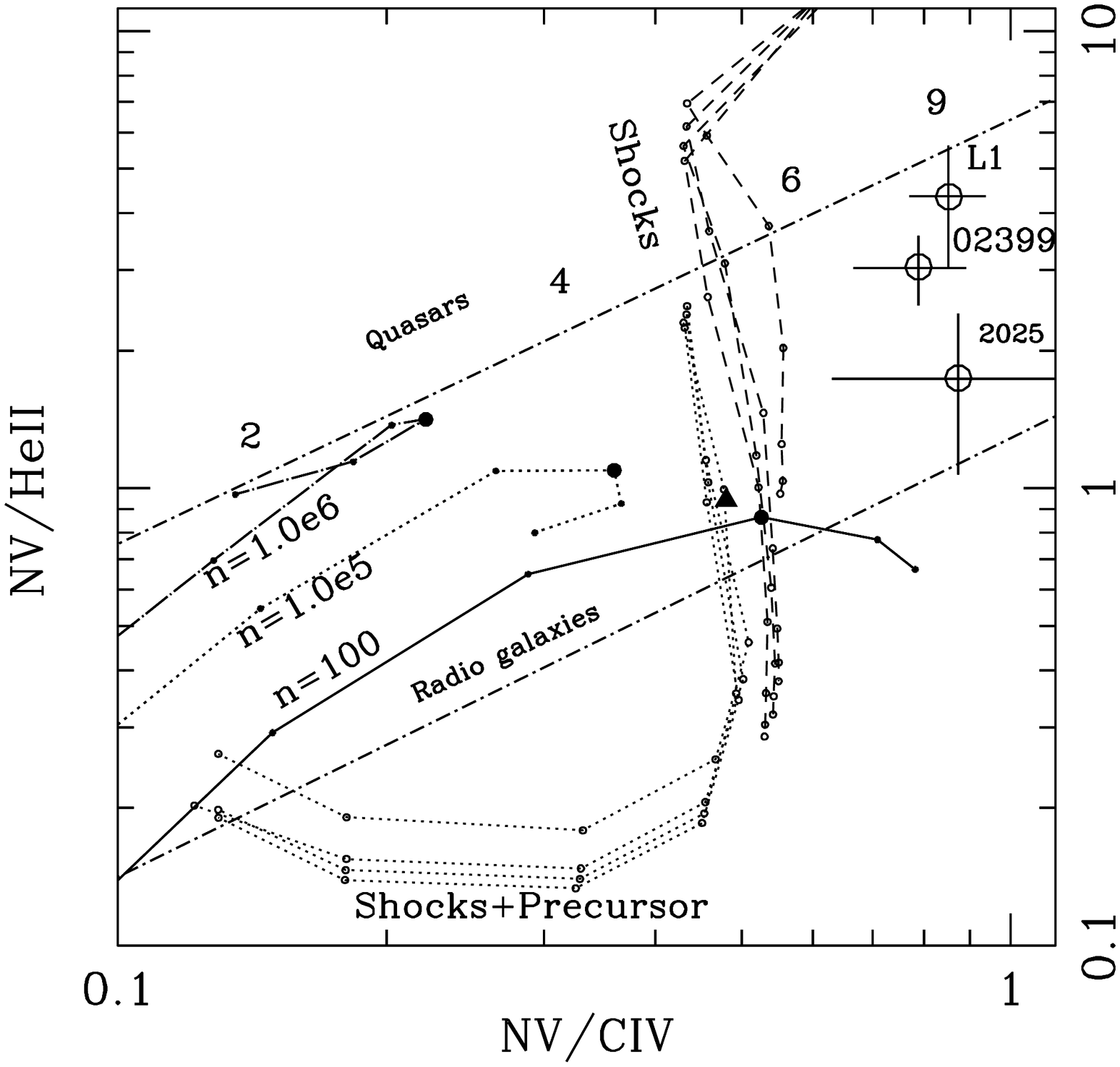}
\vspace{3in}
\caption{NV/CIV {\it vs.} NV/HeII. The same models described in Fig. \ref{Fig7}
are presented. The solid circle in each AGN sequence is the model with
U=0.1. The U=0.01 model lies outside the diagram. Therefore the models
able to explain the positions of the objects in the 
diagrams in Fig. \ref{Fig7} are unable to predict
line ratios involving the NV line;  NV is too weak.
The inclined dot-dashed lines represent
the sequence defined by high redshift quasars (Hamann \& Ferland \cite{hamm93}) and HzRG 
(Fosbury et al. \cite{fos98}, \cite{fos99}). 
The numbers on
the quasar line indicate the metallicity (in solar units) calculated by the 
models of Hamann 
and Ferland. FSC10214+4724 is indicated with the solid triangle.}
\label{Fig8}
\end{figure}

	We have also studied the NV$\lambda$1240 emission. This line was
 not often detected
 in earlier spectra of
HzRG because of limited S/N (see van Ojik \cite{ojik95}). In our small sample, only MRC2025-218 and SMM J02399-0136 have
detectable NV emission. We present in Fig. \ref{Fig8}   the diagram NV/HeII {\it vs.} NV/CIV. 
Quasars define a tight correlation in this diagram (Hamann \& Ferland \cite{hamm93}) 
which is represented as
a inclined line. Fosbury et al. (\cite{fos98},\cite{fos99}) showed 
that HzRG follow a parallel correlation to the one defined by quasars.
This is also shown in the diagram.
 We have plotted the position of
SMMJ02399-0136,  L1 and MRC2025-218.
 Interestingly, the  NV/CIV and NV/HeII line ratios measured in  these sources 
are the largest 
observed for HzRG.  MRC2025-218 lies at the top of the co\-rrelation defined by HzRG, while
 L1 lies on the  quasars correlation and also occupies the position of 
the largest  values for
the NV line ratios. Both standard AGN photoionization models (with 
solar abundances and density $n\leq$100 cm$^{-3}$) and shock models are unable to reproduce the
position
of the objects in this diagram.

\vspace{0.2cm}

\centerline{\it The emission lines spectrum of L1 in SMM J02399-0136}

\vspace{0.2cm}

	The nuclear spectrum of high $z$ quasars presents  weak or absent HeII and strong NV.
(NV/HeII$>$5 and
NV/CIV presents a large diversity with values sometimes $>$5). HeII is generally
narrower (when detected) than other lines like CIV and CIII] (Foltz et al. \cite{foltz88},
 Heckman et al. \cite{heck91}). This has been 
interpreted as the origin of an important fraction of the HeII emission ($\sim$50\%)
 in a lower velocity extranuclear 
region, possibly the ISM of the host galaxy of the quasar. This gas will have an
important contribution to the Ly$\alpha$ emission, but lines like NV, CIII] and CIV will
be  dominated by the broad line region (BLR). A view of the BLR in L1 would explain the 
weakness of HeII and the strength of NV.

However, L1 is not a quasar since the 
 lines are too narrow (FWHM$>$2500 km s$^{-1}$ in quasars). 
CIII] is broad (FWHM$\sim$6100 km s$^{-1}$),
 but this could be due to  the contamination by the SiIII]1895  doublet.
It is possible  that the lines could appear narrower and asymmetric as
the result of absorption by gas and/or dust, which could be very efficient at quenching the
resonant lines (CIV, NV, Ly$\alpha$).  However, H$\alpha$+[NII] presents also
a narrow profile ($\sim$1060 km s$^{-1}$) (IV98).
Therefore,
the emission lines are much narrower than in quasars  and  L1 is not a normal quasar. 

	The spectral properties of L1 rather suggest that it is a narrow
line active galaxy (a Seyfert 2 or narrow line quasar). Why does it
lie on the sequence defined by quasars in the NV/HeII {\it vs.} NV/CIV
diagram? Why does it  show like quasars  weak HeII and strong NV relative to the
C lines? Since the BLR 
is not visible the spectrum must be dominated
by the intermediate density na\-rrow line region ($n\sim$10$^{4-6}$) 
and/or the low density narrow line region ($n\leq$100, 
usually named the extended
emission line region, EELR). We showed above that the standard EELR models 
fail
to reproduce the observed line ratios. We present in Fig. \ref{Fig7} two  sequences
(in $U$) of models  simi\-lar to the AGN sequence described above, but for
densities $n=$10$^5$ and $n=$10$^6$.  The $n=$10$^6$ sequence
solves the discrepancies between the models and the observed line
ratios. 
 Therefore, the weakness of HeII 
can be explained if the intermediate density narrow line region
 dominates the line emission.  However, these models still fail to reproduce the
line ratios involving NV. Fig. \ref{Fig8} shows that the $n=$10$^6$ models (only
the higher ionization model is present)  predict the NV line 
too weak. On the other hand,
our models show that such high densities would produce
$\frac{[OIII]\lambda5007}{H\alpha +[NII]}\geq$4,
while the  ratio measured by IV98 is $\leq$1.
	
 One possibility is that the magnification due to the gravitational
lens is not the same for all emission lines. The magnification depends 
strongly on the source size (Trentham \cite{tren95}) being 
 higher for a smaller size of the source.
Those lines whose emission is dominant in a more nucleated region will
be more magnified than those lines preferentially emitted in a more extended
region. Since the high density  models predict stronger metal lines relative to HeII
compared to the EELR models (density $\leq$100 cm$^{-3}$) we expect CIV, CIII] and NV
 to be more nucleated and
 therefore more amplified. In this
scenario, the region emitting NV should at the same time 
be more nucleated than the regions emitting the C lines to explain the large
N/C ratios. Differential magnification has been  suggested by Lacy et al. (\cite{lacy98}) to explain the
spectroscopic properties of FSC10214+4724.

	An alternative explanation 
is that N is overabundant. NV/HeII and NV/CIV have been used both in quasars 
(Hamann \& Ferland \cite{hamm93}) and HzRG (Fosbury
et al. \cite{fos98}, \cite{fos99}) as abundance indicators.  By studying the NV/HeII {\it vs.} NV/CIV diagram
and the tight correlation defined by high redshift quasars, the authors conclude that
  the BLR of quasars at high redshift ($z>$2) present N overabundance and 
a large range in metallicities typically $\sim$1 to $\sim$10 times the solar values. 
N/C is enhanced  compared to the solar values due to secondary production
(N behaves in a different way than C and O. $N\propto Z^2$, where $Z$ is the metallicity).
 Fosbury et al.   proposed a similar explanation (but referred to the ionized gas
outside the BLR) for the correlation defined by HzRG in this diagram. 
	The failure of the solar abundance 
(both low and high density) models to explain the strength of NV (and maybe also the weakness of HeII)
suggests metal enrichment also in the ionized gas of L1. We have
calculated upper limits to the flux of the
NIII]$\lambda$1750 and NIV]$\lambda$1486 lines assuming FWHM=1600 km s$^{-1}$ (larger values are
possible. In this case we will obtain higher upper limits and the 
conclusions will not vary).  We obtain 
 $\frac{NIII]}{HeII}\leq$0.7 and  $\frac{NIV]}{HeII}\leq$0.8. These values
are well above the photoionization model predictions 
($\frac{NIII]}{HeII}\leq$0.1 and  $\frac{NIV]}{HeII}\leq$0.25). If the large
values are confirmed, the NIII] and NIV] lines will support the
interpretation of N overabundance.

\vspace{0.2cm}

\centerline{\it The emission line spectrum of  MRC2025-218}

\vspace{0.2cm}

	Fig. \ref{Fig7} shows that models with density 10$^5$ cm$^{-3}$ reproduce
the MRC2025-218 line ratios involved in these diagrams. This suggests that also in this object
the line emission is not dominated by the EELR, but by the intermediate density
narrow line region. As before, these models predict too weak NV (see Fig. \ref{Fig8}).

NV is 
 rather broad (FWHM$\sim$2800 km s$^{-1}$) compared 
to the other emission lines in this object. 
Broad NV (FWHM$\sim$3000 km s$^{-1}$) was also reported by McCarthy et al.  (1990).
This suggests that the BLR emission could contaminate the line.
The spectroscopic properties of this objects show that it is not a quasar (or BLRG).
 All the lines, including CIII] not susceptible to absorption, 
are narrow  and therefore, the emission
 is not dominated by the BLR.
 The  line ratios 
 are also 
inconsistent with quasar values
(CIV/HeII$>$7, CIII]/HeII$>$3).
 The high level of polarization of the continuum and
the fact that the line emission is not dominated by an unresolved component
(see the plateau presented by the Ly$\alpha$ in Fig. \ref{Fig3})  suggest also
that it is not a quasar.  

	However, a close look to the spectrum
 (see Fig. \ref{Fig1} top left panel and Fig. \ref{Fig5} bottom panel)
shows the presence of a very broad underlying component to the CIV line
(maybe also to HeII and CIII]) that suggests some  contribution of the BLR (scattered or direct).  
The broad NV suggests that at least this line is
contaminated by the BLR emission.  While the  fit to the CIV line to measure
the FWHM and flux of the line  neglects the broad wings seen in the spectrum, 
the contribution to NV (and CIII]) could be strong enough to broaden the line profile and
enhance the NV emission relative to the other lines. This would also explain the
anomalous strong CIII] emission.

	An alternative possibility is that MRC2025-218   is richer in metals than 
other HzRG.  
  MRC2025-218 lies at the top of the HzRG correlation. Unless we have a view on the BLR, 
 the NV diagram suggests that  MRC2025-218 is the most enriched HzRG observed.
We have
calculated upper limits to the flux of the
NIII] and NIV] lines. After taking into account the possible FWHM values
suggested by the other emission lines we obtain 
 $\frac{NIII]}{HeII}\leq$0.6 and  $\frac{NIV]}{HeII}\leq$0.8 which are
again well above the model predictions.

	In summary, the strength of NV is inconsistent in L1 and MRC2025-218
with the standard  
 model (low density and solar abundances) 
predictions. In both objects the spectrum seems to be dominated by the intermediate density 
narrow line region. This is supported by the compact appearance of the dominant line emitting
region.
This could be the case of other HzRG.
  The models suggest that N is overabundant in the ionized gas. 

\subsection{The absorption lines}

	In \S3.3 we reported the detection of several absorption features,
and the presence of PCygni profiles for CIV and SiIV lines. We discuss here the
nature of  the absorption: is it stellar or interstellar? The detection of 
stellar features would be very important;  the most convincing evidence for stars in a HzRG
has been found  in 4C41.17 (Dey et al. \cite{dey97}). On the other hand, associated 
($z_{abs}\sim z_{emis}$)
narrow absorption line systems have been found in the spectrum of several radio loud quasars
({\it e.g.} Anderson et al. \cite{and87}) and HzRG. R\"ottgering et al. (\cite{rott97}) reported the existence of deep
troughs  in the Ly$\alpha$ velocity profile of many HzRG.
 In most cases
the Ly$\alpha$ emission is absorbed over the entire spatial extent (up to 50 kpc).
The authors interpret these results as absorption by HI, physically associated with the
radio galaxy, and having a covering factor near unity. Narrow absorption troughs have been also
found in the spectral profile of the CIV line in the radio galaxy 0949-242 (R\"ottgering \& Miley \cite{rott97b}),
which is likely to be due to  associated absorption systems as well (Binette et al. \cite{bin99}).
  
The safest way to confirm  the presence of
stars is the detection of purely  stellar photospheric features, but we do not detect them in
MRC2025-218. The  identified features  may have a dominant contribution
 from interstellar absorption. 

	We have compared the absorption line
spectrum of MRC2025-218 with that of: 

	1)  4C41.17 

	2) 3C191 ($z_{emis}=$1.953), the first QSO found to have a
rich absorption line spectrum ($z_{abs}=$1.947) (Burbidge, Lynds \& Burbidge \cite{burbi66}). 
The absorption
is produced by {\it associated} absorption systems, interpreted as the consequence
of material flowing out of the nucleus of the QSO.

	3) the star forming knot B1 in the nearby starburst galaxy NGC1741 (and other
nearby starburst galaxies).

	4) star forming galaxies at $z\sim$3-4

	The EWs of the absorption features are larger than the values measured in 
4C41.17  and  NGC1741 and other starburst galaxies 
(EW$\sim$2 \AA\ , York et al. \cite{york90}). The values are consistent
(except for CIV) with the EWs in star for\-ming galaxies at $z\sim$3 
(EW$\sim$2-3.5 \AA) (Steidel et al. \cite{stei96}, Yee et al. \cite{yee96}). The agreement
is best with  the absorption lines in 3C191.
We cannot give a definitive answer on the nature of 
the absorption features in MRC2025-218.  P Cygni profiles are characteristic
of Wolf-Rayet and O star winds and have been observed in star forming galaxies
at redshift $>$3 (Steidel et al. \cite{stei96}). 
However, the features we detect are
 highly contaminated
by interstellar absorption in normal starburst galaxies. 
P Cygni  profiles have been  observed  both  in high and low
ionization lines in some HzRG (see Fig.~5 in Dey \cite{dey98}). The author
suggests that the absorption is produced by fast outflowing material moving
at high velocity relative to the galaxy (this could not explain redshifted
 absorption
features observed in some HzRG, though [R\"ottgering et al. \cite{rott97}]).

\section{Summary and conclusions}

	We have studied the UV spectra of 3 high
redshift radio galaxies and the hyperluminous system SMM J02399-0136.
All objects present extended continuum and line emission along the
radio axis. Several spatial components are found in all objects.

	The line ratios of the active galaxy (L1) in the system  SMM J02399-0136
and the radio galaxy MRC2025-218 suggest that the emission line spectra are dominated
by the intermediate density narrow line region ($n\sim$10$^{5-6}$ cm$^{-3}$),
 rather than the
low density more  extended gas ($n\leq$100 cm$^{-3}$). This could also be the case
for other HzRG.

We find that MRC2025-218 and L1
show unusually strong NV, inconsistent with solar abundance model predictions.
Comparison with studies of high redshift quasars and radio
galaxies suggest that N is overabundant in both objects.  
An alternative
possibility for L1 is that NV is emitted in a more nucleated region and
is more amplified by the gravitational lens. An alternative possibility
for MRC2025-218 is that emission from the broad line region contaminates
the NV line. 

	We detect several absorption features in the continuum of
MRC2025-218. No pure photospheric features have been identified
unambiguously. We observe PCygni profiles in some of the lines.
The nature of the absorption is not clear. It could be due to
stars or to associated absorption systems, as observed in other
HzRG.

\begin{acknowledgements}
This work is based on spectroscopic data obtained at La Silla Observatory.
Thanks to Bob Goodrich and Marshall Cohen for providing the spectrum of FSC10214+4724
and Joel Vernet for the average HzRG spectrum. 
M. Villar-Mart\'\i n thanks Kirsty Green and Jane Weir for useful
discussions.
\end{acknowledgements}


\begin{thebibliography}{}


\bibitem[1987]{and87} Anderson S.F, Weymann R.J., Foltz C.B., Chaffee F.,
 1987, AJ 94, 278

\bibitem[1967]{bahc67} Bahcall J., Sargent W., Schmidt M. , 1967, ApJ, 149, 11

\bibitem[1999]{bin99} Binette  L., Kurk J., Villar-Mart\'\i n M., R\"ottgering H., Hunstead R., 1999, A\&A,
submitted

\bibitem[1966]{burbi66} Burbidge E., Lynds C., Burbidge M., 1966, ApJ, 144, 447

\bibitem[1984]{burs84} Burstein D., Heiles C., 1984, ApJS, 54, 33

\bibitem[1998]{best98} Best P., Longair M., R\"ottgering H., 1998, MNRAS, 295, 549

\bibitem[1999]{borne99} Borne K., Bushouse H., Colina L., Lucas R., 1999, in {\it 
After the dark ages: When galaxies were young (the universe at 2$<z<$5)},
Conf. Procc. Holt S. \& Smith E. eds. American Institute of Physics Press,
p220

\bibitem[1989]{card89} Cardelli J., Clayton G., Mathis J., 1989, ApJ, 345, 245

\bibitem[1987]{chamb87} Chambers K.C., Miley G.K., van Breugel W., 1987, Nature 329, 604  

\bibitem[1996]{cim94} Cimatti A., di Serego Alighieri S., Field G.,
Fosbury R., 1994, ApJ, 422, 562

\bibitem[1996]{cim96} Cimatti A., Dey A., van Breugel W., Antonucci R., Spinrad H., 1996, ApJ, 465, 145

\bibitem[1997]{cim97} Cimatti A., Dey A., van Breugel W., Hurt T., Antonucci R., 1997, ApJ, 476, 677

\bibitem[1998]{cim98} Cimatti A., di Serego Alighieri S., Vernet J., Cohen M., Fosbury R., 1998, ApJ 499, 21

\bibitem[1996]{conti96} Conti P., Leitherer C., Vacca W., 1996, ApJ, 461, 87

\bibitem[1997]{dey97} Dey A., van Breugel W., Vacca W., Antonucci R., 1997, ApJ, 490, 698   

\bibitem[1998]{dey98} Dey A., 1998, in {\it The
most distant radio galaxies}, Amsterdam, The Netherlands, October 1997.
 R\"ottgering, Best \& Lehnert eds. (astro-ph/9803137)



\bibitem[1995]{dickson95} Dickson R., Tadhunter C., Shaw M., Clark N., Morganti R., 1995,
 MNRAS, 273, 29


\bibitem[1996]{dop96} Dopita M.A., Sutherland R.S., 1995, ApJ, 455, 468

\bibitem[1997]{ferr97} Ferruit P., Binette L., Sutherland R., P\'econtal E., A\&A, 322, 73

\bibitem[1988]{foltz88} Foltz C., Chaffee F., Weymann T., Anderson S., 1988, in {\it QSO Absorption
Lines: Probing the Universe}, ed. J.C. Blades, D. Turnshek \& C. Norman (Cambrdige: Cambridge University
Press), 53

\bibitem[1998]{fos98}Fosbury R., Vernet J., Villar-Mart\'\i n M.,  Cohen M., Cimatti A., 
di Serego Alighieri S., McCarthy P., in {\it NICMOS and the VLT: A New Era of High Resolution Near Infrared
Imaging and Spectroscopy}, 1998, Pula, Sardinia, Italy, 26-27 June,  1998, ESO
Conference and Workshop Proceedings 55, p. 190.  Wolfram Freudling and Richard
Hook eds.

\bibitem[1999]{fos99}Fosbury R., Vernet J., Villar-Mart\'\i n M.,  Cohen M., Cimatti A., 
di Serego Alighieri S., McCarthy P., 1999, in {\it ESO Conference on Chemical Evolution from Zero
to High Redshift}, Garching, Germany, October 14-16 1998. ESO Astrophysics
Symposia, Eds. J. Walsh and M. Rosa, Springer

\bibitem[1998]{goo96} Goodrich R., Miller J., Martel A., Cohen M., Tran H., Ogle P.,
 Vermelulen R., 1996, ApJ, 456, 9

\bibitem[1993]{hamm93} Hamann E., Ferland G., 1993, ApJ, 418, 11 


\bibitem[1991]{heck91} Heckman T., Lehnert M., Miley G., van Breugel W., 1991, ApJ, 381, 373


\bibitem[1998]{iv98}Ivison R., Smail I., Le Borgne J., Blain A., Kneib J.,
B\'ezecourt J., Kerr T., Davies J.,  1998, MNRAS, 298, 593, IV98

\bibitem[1999]{iv99} Ivison R., Smail I., Blain A., Kneib J., Frayer D.,
1999, to appear in the proccedings of the 1998 Ringberg workshop of 
ultraluminous galaxies (also astro-ph/9901361)

\bibitem[1992]{kor92} Kormendy J., Sanders B., 1992, ApJ, 390, 53

\bibitem[1998]{lacy98} Lacy M., Rawlings S., Serjeant S., 1998, MNRAS, 299, 1220

\bibitem[1983]{lilly83} Lilly S., Longair M., McLean I., 1983, Nature, 301, 488

\bibitem[1987]{mac87}  McCarthy P., Spinrad H., Djorgovsky S.,
Strauss M.A., van Breugel W., Liebert J., 1987, ApJ, 319, L39

\bibitem[1990]{mac90}  McCarthy P., Kapahi V., van Breugel W., Subrahmanya C., 1990, 
AJ 100, 1014

\bibitem[1992]{mac92}  McCarthy P.,  Elston R., Eisenhardt P., 1992,
ApJ, 387, 29

\bibitem[1996]{mac96}McCarthy P.J., Baum S., Spinrad H., 1996, ApJS
106, 281

\bibitem[1999]{pente99}Pentericci L., R\"ottgering H., Miley G., McCarthy P., Spinrad H., van
Breugel W., Macchetto F., 1999, A\&A, 341, 329

\bibitem[1989]{rees89}Rees M., 1989, MNRAS, 239, 1

\bibitem[1994]{rott94} R\"ottgering H., Lacy M., Miley G., Chambers K., Saunders R.,
1994, A\&ASS, 108, 79

\bibitem[1995]{rott95}R\"ottgering H., Miley G., Chambers K., Macchetto F., 1995
A\&ASS, 114, 51

\bibitem[1997]{rott97}R\"ottgering H., van Ojik R., Miley G., Chambers K.,
van Breugel W., de Koff S., 1997, A\&A, 326 505

\bibitem[1997]{rott97b} R\"ottgering H., Miley G., in {\it The Early Universe with the VLT}, 
Bergeron J. ed., Springer: Berlin, p285

\bibitem[1988]{sand88} Sanders D., Soiffer B., Elias J., Madore B.,
Matthews K., Neugebauer G., Scoville N., 1988, ApJ, 325, 74

\bibitem[1996]{sand96} Sanders D., Mirabel I., 1996, ARA\&A, 34, 749

\bibitem[1996]{stei96} Steidel C., Giavalisco M., Pettini M., Dickinson M., 
Adelberguer K., 1996, ApJL, 462, L17

\bibitem[1989]{tad89} Tadhunter C., Fosbury R., di Serego Alighieri S.,
1989, in {\it BL Lac Objects} Conf. Procc., Maraschi L., Maccacaro T. \&
Ulrich M.H. eds. Springer Verlag, Berlin, p. 79

\bibitem[1995]{tren95} Trentham N., 1995, MNRAS, 277, 616 

\bibitem[1995]{ojik95}van Ojik R., 1995, Ph.D. Thesis, University of Leiden

\bibitem[1999]{breu99}van Breugel W., Stanford S., Spinrad H, Stern D., 
Graham J.,
1998, ApJ, 502, 614
\bibitem[1999]{ver99} Vernet J., Fosbury R., Villar-Mart\'\i n M., CohenM., Cimatti A., 
di Serego Alighieri S., 1999, in {\it ESO Conference on Chemical Evolution from Zero
to High Redshift}, Garching, Germany, October 14-16 1998. ESO Astrophysics
Symposia, Eds. J. Walsh and M. Rosa, Springer


\bibitem[1997]{vill97}Villar-Mart\'\i n M., Tadhunter C., Clark N.,
 1997, A\&A, 323, 21

\bibitem[1999]{vill99}Villar-Mart\'\i n M., Binette L., Fosbury R.A.E., 
1999, A\&A, 346, 7 (VMB99)

\bibitem[1996]{yee96} Yee H., Ellingson E., Bechtold J., Carlberg R., Cuillandre J.,
 1996, AJ, 111, 1783

\bibitem[1990]{york90} York D., Caulet A., Rybsky P., Allagher J., Blades J.,
Morton D., Wamsteker W., 1990, ApJ, 353, 413

\end{thebibliography}
\end{document}